\documentclass[12pt]{article}

\usepackage{amsmath,amssymb}
\usepackage{graphicx}
\usepackage{caption}
\usepackage{color}
\usepackage{dcolumn}
\usepackage{bm}
\usepackage[numbers,super,comma,sort&compress]{natbib}
\usepackage[version=3]{mhchem}
\usepackage{booktabs}
\usepackage{siunitx}
\usepackage{color}
\usepackage{placeins}

\let\Oldsection\section
\renewcommand{\section}{\FloatBarrier\Oldsection}

\let\Oldsubsection\subsection
\renewcommand{\subsection}{\FloatBarrier\Oldsubsection}

\let\Oldsubsubsection\subsubsection
\renewcommand{\subsubsection}{\FloatBarrier\Oldsubsubsection}

\definecolor{background-color}{gray}{0.98}

\title{Performance of SCAN density functional for a set of ionic liquid ion pairs}

\author{Karl Karu\thanks{Institute of Chemistry, University of Tartu, Ravila 14a, Tartu 50411, Estonia}, Maksim Mi\v{s}in\footnotemark[1], Heigo Ers\footnotemark[1], Jianwei Sun\thanks{Department of Physics and Engineering Physics, Tulane University, New Orleans, USA 70118}, Vladislav Ivani\v{s}t\v{s}ev\footnotemark[1]}

\begin{document}

\maketitle
\noindent\textbf{This is the pre-peer reviewed version of the article entitled "Performance of SCAN density functional for a set of ionic liquid ion pairs", which is accepted to be published in the International Journal of Quantum Chemistry in final form at DOI:10.1002/qua.25582. This article may be used for non-commercial purposes in accordance with Wiley Terms and Conditions for Self-Archiving.}
\begin{abstract}
Computational chemistry is a powerful tool for the discovery of novel materials. In particular, it is used to simulate ionic liquids in search of electrolytes for electrochemical applications. Herein, the choice of the computational method is not trivial, as it has to be both efficient and accurate. Density functional theory (DFT) methods with appropriate corrections for the systematic weaknesses can give precision close to that of the post-Hartree--Fock coupled cluster methods with a fraction of their cost. Thence, we have evaluated the performance of a recently developed non-empirical Strongly Constrained and Appropriately Normed (SCAN) density functional on electronic structure calculations of ionic liquid ion pairs. The performance of SCAN and other popular functionals (PBE, M06-L, B2PLYP) among with Grimme's dispersion correction and Boys--Bernardi basis set superposition error correction was compared to DLPNO-CCSD(T)/CBS. We show that SCAN reproduces coupled-cluster results for describing the employed dataset of 48 ion pairs.
\\[10mm]
\end{abstract}

\clearpage


  \makeatletter
  \renewcommand\@biblabel[1]{#1.}
  \makeatother

\bibliographystyle{apsrev}

\renewcommand{\baselinestretch}{1.5}
\normalsize

\clearpage

\section*{\sffamily \Large INTRODUCTION} 

Ionic liquids are promising solvents that have been extensively studied over the last few decades. Their tunable properties make them advantageous candidates for various electrochemical applications.\cite{Macfarlane2014ugy,Fedorov2014uyy} The high price of commercially available ionic liquids remains an obstacle for an extensive use of ionic liquids.\cite{Plechkova2008tnl} However, this is also a strong stimulus for the search of advanced ionic liquids.\cite{Macfarlane2014ugy}

A useful toolkit for studying ionic liquids includes various computer simulation methods.\cite{Izgorodina2017tmr} Simulations have been growing in popularity, hand in hand with an exponential increase in the computational capabilities.\cite{Salanne2015vnn,Izgorodina2011usg} For instance, computational screening allows estimating the properties of many candidates even prior to their synthesis.\cite{Zvereva2010uay,Katsyuba2007tfv,Cheng2015tmj,Husch2015vnl,Firaha2015wvi} Therefore it helps to determine, which candidates are the best for a given application. Also, molecular dynamics simulations provide an insight into both structure and dynamics of specific ionic liquids both in bulk and near various interfaces.\cite{Salanne2015vnn,Maginn2009wla,Ivanistsev2015usp,Xu2015ues,Docampo-alvarez2016utk,Skarmoutsos2012udr} However, they require careful parametrisation of the force fields used.\cite{Salanne2015vnn,Maginn2009wla,Borodin2009thn,Chaban2011uhg,Bernardes2016vwr,Hunt2006uip} On the contrary, without parametrisation, quantum mechanical calculations allow rapid exploration of the space of ionic liquids, which is vast due to numerous possible anion--cation combinations.\cite{Izgorodina2017tmr,Karu2016tki} 

Fast computational methods are essential for both high-throughput screening and large-size simulations. At the same time, the methods have to be accurate enough to capture all of the physical interactions within the ionic liquids. The density functional theory (DFT) methods offer a good trade-off between speed and accuracy.\cite{Izgorodina2017tmr,Zahn2013uif,Izgorodina2009ubs} The success of DFT depends whether employed exchange--correlation functional can adequately describe the system of interest.\cite{Lage-estebanez2017vki} In practice, the necessity for faster and scalable methods attracts the attention towards pure DFT functionals and prevents the use of generally more accurate yet more expensive hybrid functionals.

\begin{figure}[htbp]   
\begin{center}
  \includegraphics[width=6.5in]{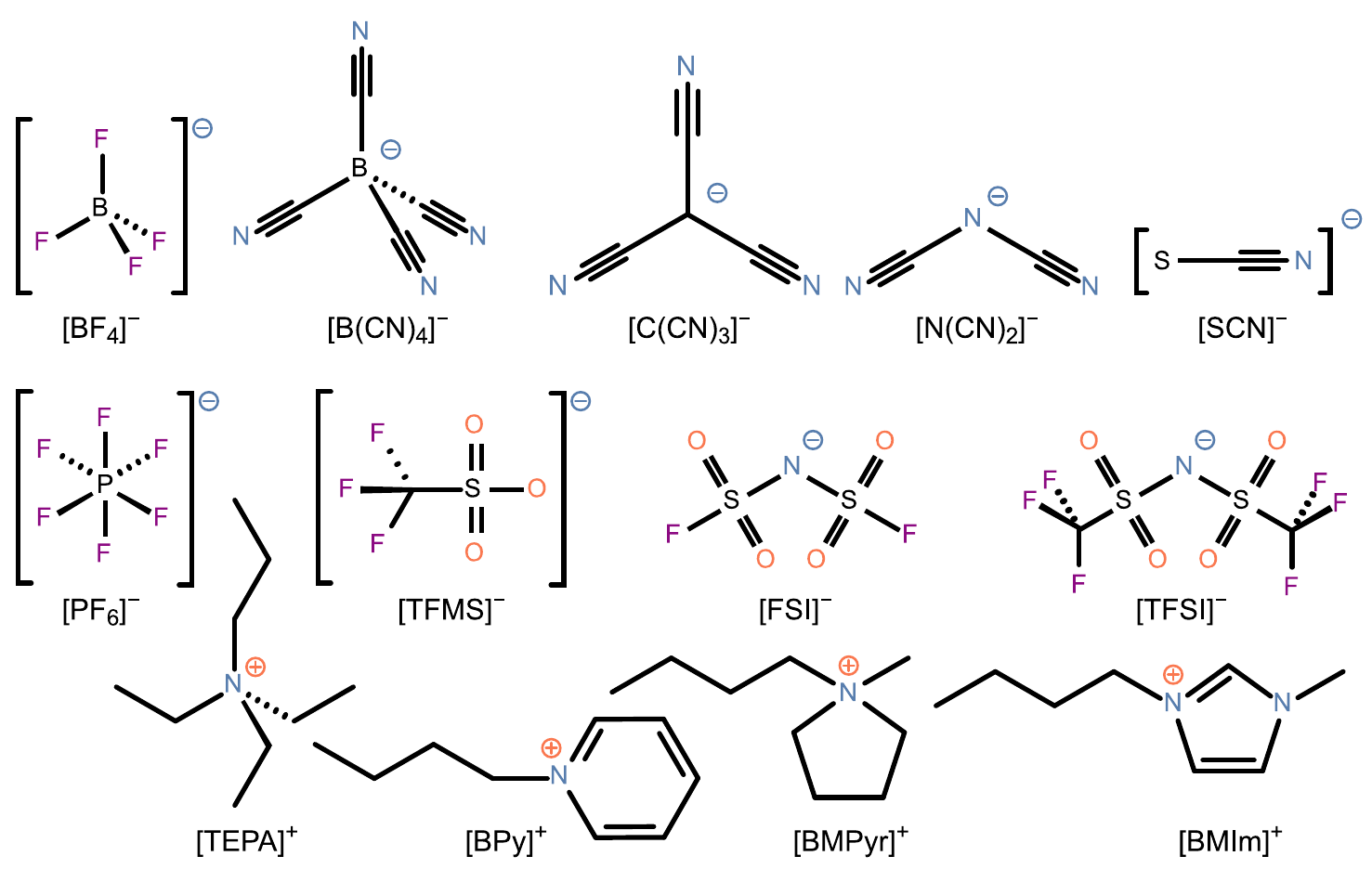}
  \caption{2D structural formulas of the anions and the cations used to combine ion pairs. From the top left: tetrafluoroborate, tetracyanoborate, tricyanomethanide, dicyanamide, isothiocyanate, hexafluorophosphate, trifluoromethylsulfonate, bis(fluorosulfonyl)imide, bis[(trifluoromethyl)sulfonyl]imide, N,N,N-triethyl-N-propylammonium, 1-butylpyridinium, 1-butyl-1-methylpyrrolidinium, 1-butyl-3-methylimidazolium. Chloride, bromide and iodide ions were also employed but are not shown in the Figure.}
  \label{fig:2Dions}
\end{center}
\end{figure}

The above makes the recently proposed strongly constrained and appropriately normed (SCAN) functional an intriguing candidate for investigations of ionic liquids. It is non-hybrid and therefore significantly faster than a hybrid functional,\cite{Sun2015whb} while in many cases it is even more accurate.\cite{Sun2016tdn} To evaluate the performance of SCAN on ionic liquids, we made a dataset of 48 ion pairs by combining 4 cations with 12 anions (Figure~\ref{fig:2Dions}).

The aim of the current work is twofold. Firstly, to verify that the SCAN functional is suitable for the energy and dipole moment calculations of ionic liquid ion pairs, and to compare its predictions to other commonly used functionals. Secondly, to provide a dataset of ion pair geometries and energies, suitable for testing computational methods.

\section*{\sffamily \Large METHODOLOGY}

\subsection*{Automatic workflow}
\begin{figure}[htbp]
\begin{center}
  \includegraphics[width=3.25in]{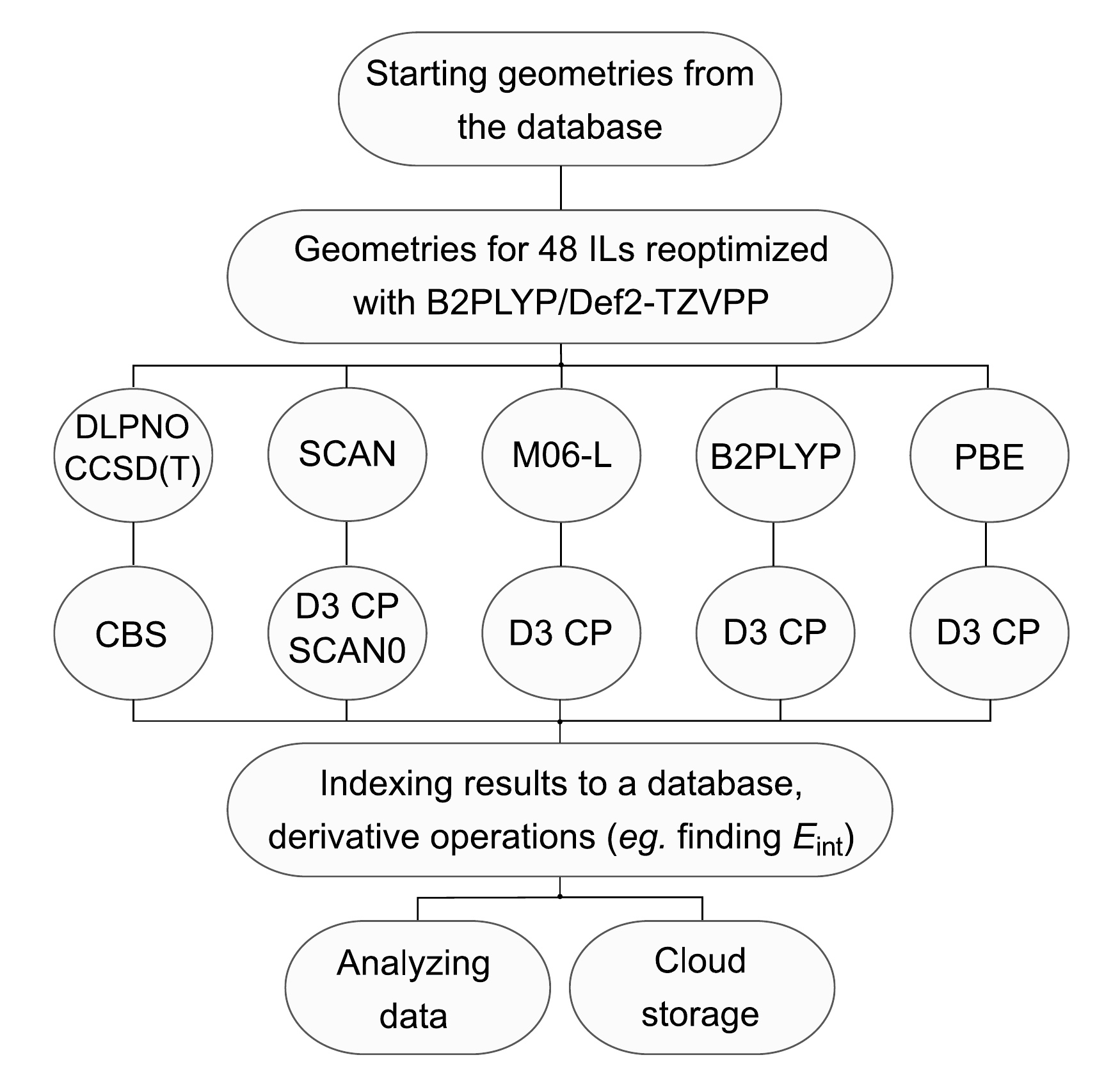}
  \caption{A graphical scheme of the constructed NaRIBaS workflow.}
  \label{fig:workflow}
\end{center}
\end{figure}

We employed the scripting framework NaRIBaS (Nanomaterials and Room-temperature Ionic liquids in Bulk and Slab) for repetitive calculations and data analysis.\cite{Ivanistsev2015tvg} Namely, as shown in Figure~\ref{fig:workflow}, the NaRIBaS workflow was used to optimise the geometries of the ion pairs, to prepare inputs, to carry out single point calculations, to gather the results into a simple \textit{json}-format database, and to perform the analysis. Herewith, the SCAN and SCAN0 calculations were performed using a modified version of Gaussian~03 code,\cite{Frisch2008wnt}  while all other calculations were run using the Orca~3.0.3 package.\cite{Neese2012vzl}

\subsection*{Reference method and the development of the dataset}

For the geometry optimisation, we employed the double-hybrid B2PLYP functional, in which the DFT exchange energy is corrected with the exact Hartree--Fock exchange, and the correlation energy with that of second-order M{\o}ller--Plesset (MP2) perturbation theory.\cite{Grimme2006ufy} See section 1.5 of the Supporting Information (SI) for details. 

The geometry optimisation was done with Ahlrichs' type triple-$\zeta$ Def2-TZVPP basis set with polarization functions on all atoms.\cite{Schafer1992wfc,Weigend2005wsp} Tight self-consistent field convergence parameters, SCF integration grid~5 with final grid~6 and tight geometry optimisation grids as defined in Orca~3.0.3 were used. The resolution of identity approximation was employed to speed up the calculations with approximations for Coulomb integrals, numerical Hartree--Fock exchange integrals, and MP2 correlation integrals. Grimme's dispersion correction with Becke--Johnson damping was also employed.\cite{Grimme2010vgm,Johnson2006toc,Johnson2005vtc} Most of the starting geometries were obtained from the Ref. \citenum{Ivanistsev2015tvg} GitHub repository. 

As the reference we used an approximation to the coupled cluster method with singles, doubles and perturbative triples excitations (CCSD(T)) paired with triple-$\zeta$ basis sets (Def2-TZVPP), and extrapolated to the complete basis set limit (CBS):

\begin{itemize}
\item To manage the size-scaling of CCSD(T), we employed the domain-based local pair natural orbital (DLPNO) approximation with tight parameters ($T_{\text{CutPairs}}= 10^{-5}, T_{\text{CutPNO}}=10^{-7}, T_{\text{CutMKN}}=10^{-4}$).\footnote{According to Ref. \citenum{Liakos2015wgp}, DLPNO-CCSD(T) with tight parameters has less than \SI{1}{kJ\per\mole} standard deviation from CCSD(T) for the FH dataset,\cite{Friedrich2013wgp} S66 database\cite{Rezac2011wuo} and two datasets containing conformational energies
of butane-1,4-diol\cite{Jesus2008ufy} and melatonin.\cite{Fogueri2013wnf}}
\item Spin-component scaled second-order M{\o}ller--Plesset perturbation theory (SCS-MP2)\cite{Grimme2003ttx} calculations were run with both triple-$\zeta$ and quadruple-$\zeta$ split-valence basis sets (Def2-TZVPP and Def2-QZVPP), and resolution of the identity approximation.\footnote{According to Ref. \citenum{Friedrich2015vlq} a similar approach, which includes the DLPNO approximation and uses MP2 results to extrapolate the DLPNO-CCSD(T) energies, has less than \SI{2}{kJ\per\mole} mean deviation and less than \SI{3}{kJ\per\mole} mean average deviation from the honestly extrapolated pure CCSD(T).} Effective core potentials for iodine on the Def2-QZVPP level were employed.\cite{Peterson2003wxh} The SCS-MP2 energies were used to extrapolate the DLPNO-CCSD(T) energies to CBS.
\end{itemize}

DLPNO-CCSD(T), SCS-MP2 and the employed extrapolation method are discussed in sections 1.1--1.3 of the SI. 

\subsection*{\sffamily \large Non-hybrid density functionals}
Even with the rapid expansion of computational power, the couple-clusters and related methods are still too demanding for the computational screening of ionic liquids. Presently, a promising group of methods for this task is the so-called pure (non-hybrid) DFT approaches that do not incorporate Hartree--Fock or MP2 calculations.\cite{Zhao2004vax} 


An intriguing new pure DFT functional is the strongly constrained and appropriately normed (SCAN) functional by Sun \textit{et al.}\cite{Sun2015whb} SCAN incorporates seventeen exact constraints applicable for a semi-local functional, including the tight lower bound on the exchange energy.\cite{Perdew2014web,Sun2015whb} Unlike many other popular approaches (such as M06-2X, B3LYP) the functional is not based on the experimentally measured quantities; this makes SCAN potentially suitable for materials that have not been previously characterised. Additionally, the SCAN functional turns out to be superior to the classical and widely employed Perdew--Burke--Ernzerhof (PBE) functional in describing molecular compounds and solids.\cite{Sun2015whb} It can accurately predict geometries and energies of diversely bonded molecules and materials.\cite{Sun2016tdn}

To contrast the performance of SCAN we picked two popular non-hybrid functionals: PBE\cite{Perdew1996weg} and M06-L.\cite{Zhao2006udj} Generally, M06-L can describe dispersion interactions quite well; however, PBE tends to underestimate their strength.\cite{Izgorodina2009ubs,Marom2011udg} To overcome this difficulty we employed Grimme's dispersion corrections combined with Boys--Bernardi counterpoise correction (CP) that accounts for basis set superposition error.\cite{Grimme2010vgm,Boys1970vzj} These corrections were shown to markedly improve the accuracy of a wide variety of functionals including SCAN.\cite{Zahn2013uif,Grimme2012vux,Brandenburg2016trw}


All single-point calculation were performed on the final structures obtained from the optimised B2PLYP geometries. The single-point PBE, M06-L, and B2PLYP calculations were conducted using the Def2-TZVPP basis set, tight SCF convergence parameters, SCF integration grid~7, and the resolution of the identity approximation for Coulomb integrals as defined in Orca 3.0.3. For single-point B2PLYP calculations, the same RI-approximations were used as for the optimisation. The Grimme's D3 correction was added to the final energies with Becke--Johnson damping.\cite{Grimme2010vgm,Johnson2006toc,Johnson2005vtc} Additionally, the CP correction procedure was used to account for the basis set superposition error.\cite{Boys1970vzj} For details on the corrections see sections 1.6 and 1.7 of the SI .

The SCAN calculations, as they were run with the developer's version of Gaussian~03 code, employed 6-311++G(3df,3pd) basis set instead of Def2-TZVPP. Herewith for iodine, the diffuse functions were not included.\cite{Mclean1980wgs,Krishnan1980wqv} Tight SCF convergence parameters and ultrafine SCF integration grids were employed as defined in Gaussian 03. Additionally, we calculated the D3 correction with parameters from Ref. \citenum{Brandenburg2016trw} as well as with rescaled parameters denoted as D3*, and the Gaussian~03 built-in Boys--Bernardi CP correction. Rescaling the D3 parameters is described in the section 1.8 of the SI. For selected ion pairs, the hybrid version of SCAN with $25\%$ of Hartree--Fock exchange (SCAN0), discussed in SI (section 1.4), was also employed on the same basis set.\cite{Hui2016ust}

\subsection*{\sffamily \large Analysis}

The main metrics for evaluating the performance of the functionals in this work are values of interaction energies and dipole moments. The interaction energy of an ion pair is defined as:
\begin{equation}\label{eq:eint}
E_{\text{int}}=E_{\text{ion pair}}-(E_{\text{cation}}+E_{\text{anion}}),
\end{equation}
where $E_{\text{ion pair}}$ denotes the electronic energy of the ion pair, $E_{\text{cation}}$ the electronic energy of the cation and $E_{\text{anion}}$ the electronic energy of the anion calculated at the optimised pair geometry. Interaction energies calculated with the DFT functionals were compared against the corresponding values obtained with the reference method -- DLPNO-CCSD(T) extrapolated to CBS. 

To evaluate the performance of the functionals we used the following statistical parameters: maximum deviation (MAXD), mean absolute deviation (MAD), mean deviation or bias (MD), standard deviation of error of predictions (SDEP), and correlation coefficient ($r$):
\begin{align}
\label{eq:error_metrics}
\mathrm{MAXD} = \max |D|, && \mathrm{MAD} = \sum^{N}_{i=1} \frac{|D_i|}{N}\nonumber, &&
\mathrm{MD} = \sum^{N}_{i=1} \frac{ D_i}{N}, \\
\mathrm{SDEP} = \sqrt{\frac{1}{N}\sum^{N}_{i=1} (D_i  - \bar{D_i})^2 }, && r= \frac{\mathrm{cov}(Q_i, Q_i^{\mathrm{Ref.}})}{\sigma(Q) \sigma(Q^{\mathrm{Ref.}})}, &&
\end{align}
where $N$ is the number of all ion pairs in the dataset, $D_i = Q_i -Q_i^{\text{Ref.}}$ stands for deviation, $Q_i$ denotes a calculated quantity, while $Q_i^{\text{Ref.}}$ represents the same quantity calculated using the reference method. In the equation defining correlation coefficient $r$, $\text{cov}$ is the co-variance between predicted and reference values, and $\sigma$ stands for the standard deviation of the values.

The box-plot format is used for the presentation of results in Figures \ref{fig:deviations}, \ref{fig:hal} and \ref{fig:nohal}. In a box-plot the first and the third quartiles of the given dataset are represented by lower and upper box edges, the second quartile (\textit{i.e.} median) is represented by a horizontal line within a box, and the whiskers extend to the minimum and the maximum values. Outliers beyond 1.5 interquartile range of the box are portrayed separately, in which case the corresponding whisker is limited to $\pm{1.5}$ interquartile range beyond the box.


\section*{\sffamily \Large RESULTS AND DISCUSSION}

\subsection*{\sffamily \large The dataset of ionic liquid ion pairs}

The dataset is composed of commonly used ionic liquid cations and anions. The list of anions includes halide, cyanide, borate, sulphonate, phosphate, and imide-based anions (Figure~\ref{fig:2Dions}).

After the geometry optimisation, the cation and anion in each ion pair ended up relatively close to each other. The distances between the geometric centers of the two are in the range between \SI{2.7}{\angstrom} in the case of 1-butyl-3-methylimidazolium chloride, and \SI{5.1}{\angstrom} for N,N,N-triethyl-N-propylammonium bis[(trifluoromethyl)sulfonyl]imide. The majority of the ion pairs has dipole moment around \SI{5}{D}. The smallest dipole moment of \SI{2.8}{D} has 1-butylpyridinium chloride, while the largest dipole moment of \SI{7.3}{D} belongs to N,N,N-triethyl-N-propylammonium tetracyanoborate. The majority of ion pairs has the interaction energy between \SI{-380}{kJ\per\mole} and \SI{-340}{kJ\per\mole}. The most weakly associated ion pair in the dataset is N,N,N-triethyl-N-propylammonium tetracyanoborate ($E_\mathrm{{int}}=\SI{-291}{kJ\per\mole}$), while the most strongly bound ion pair is 1-butylpyridinium chloride ($E_\mathrm{{int}}=\SI{-410}{kJ\per\mole}$). The ion pair interaction energy characterises the cohesion of ionic liquids. It is related to properties such as viscosity, diffusion coefficients, and surface tension.\cite{Karu2016tki,Bernard2010vhu,Borodin2009wdo} It also serves as an attractive benchmark metrics, since it is reasonable to suggest that DFT functionals capable of predicting interaction energies of ion pairs will be able to accurately predict the cohesion in bulk ionic liquids and at interfaces.

Distributions of distances between the ions, dipole moments, and interaction energies are provided in the SI (Figure S2, Table S1). The whole dataset along with the optimised geometries is available at Ref. \citenum{Karu2017ttk}) GitHub repository and can be used for evaluating the performance of other computational methods. 
\subsection*{\sffamily \large Overall performance of the tested functionals}

\begin{figure}[htbp]
\begin{center}
  \includegraphics[width=3.25in]{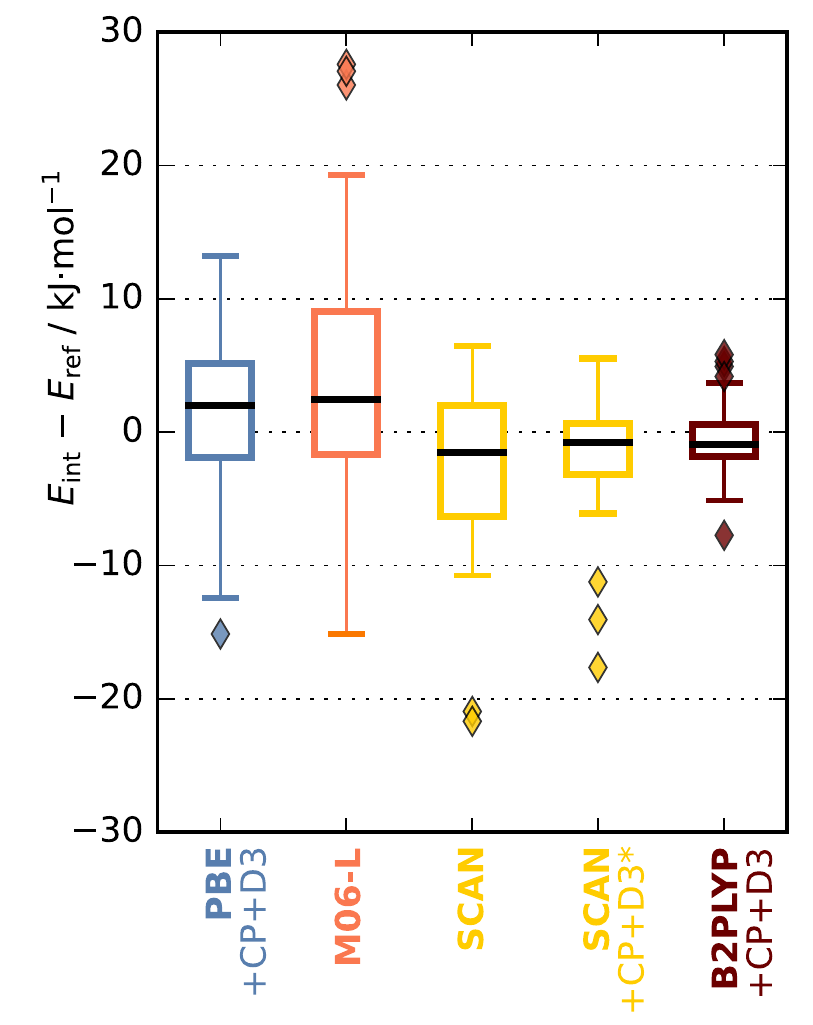}
  \caption{The distribution of errors in interaction energies (relative to the reference method) for the employed DFT methods with the most relevant corrections.}
  \label{fig:deviations}
  \end{center}
\end{figure}

The performance of the tested functionals for estimating the interaction energies is shown in Figure~\ref{fig:deviations}. For the sake of clarity, we show only the best performing approach(es) for each functional. Interaction energies obtained with B2PLYP are very close to those of the reference method and surpass all other employed methods in accuracy. B2PLYP, however, contains contributions from MP2, so it is also considerably slower than the other methods. Furthermore, the fact that the geometries were optimised on the B2PLYP/Def2-TZVPP level might also skew the results in its favour.

The other methods presented in Figure~\ref{fig:deviations} also have smaller than \SI{5}{kJ\per\mole} median error, but the distributions of errors vary considerably across the methods. One can see that errors within SCAN+CP+D3* functional are quite systematic, with the majority falling into \num{1} to \SI{-3}{kJ\per\mole} range. On the other hand, other functionals show much larger variations, with the most dramatic case, M06-L, having errors that vary between \num{20} and \SI{-10}{kJ\per\mole}.

It is worth separately addressing the effects of the CP and D3 corrections. The CP corrects the basis set superposition error (typically positive), while the D3 improves the inadequate description of dispersion interactions (typically negative). These errors are common to the majority of the DFT methods. The functionals performance with and without the different corrections is demonstrated in Table~\ref{tab:deviats}. When CP and D3 are applied, they improve the mean absolute deviation (MAD) of both PBE and B2PLYP, as well as decrease the standard deviation of error of prediction (SDEP) in the case of SCAN. Using D3* with rescaled parameters further decreases the deviations of CP-corrected SCAN, which is discussed in the next section.


\begin{table}[htbp]
  \centering
  \begin{tabular}{llrrrr}
 \toprule
Functional & Basis set & MAXD & MAD  & MD & SDEP  \\
 &  & \SI{}{kJ\per\mole} & \SI{}{kJ\per\mole} & \SI{}{kJ\per\mole} & \SI{}{kJ\per\mole}\\ 
\midrule
\midrule
PBE&Def2-TZVPP&32.8&17.0&$14.6$&$12.0$ \\
PBE+D3&Def2-TZVPP&32.5&7.0&$-6.5$&$7.0$ \\
PBE+CP&Def2-TZVPP&40.1&22.8&$22.6$&$9.3$ \\
PBE+D3+CP&Def2-TZVPP&\textbf{15.1}&\textbf{4.7}&$\textbf{1.6}$&\textbf{5.8} \\
\midrule
M06-L&Def2-TZVPP&27.6&\textbf{7.6}&\textbf{3.7}&\textbf{9.5} \\
M06-L+D3&Def2-TZVPP&85.3&31.3&$-20.6$&$26.8$ \\
M06-L+CP&Def2-TZVPP&75.7&33.7&$-25.5$&$25.3$ \\
M06-L+D3+CP&Def2-TZVPP&\textbf{23.1}&10.2&$4.8$&$10.3$ \\
\midrule
SCAN&6-311++G(3df,3pd)&21.7&4.9&$-2.4$&$6.2$ \\
SCAN+D3&6-311++G(3df,3pd)&25.3&10.4&$-10.1$&$5.3$ \\
SCAN+D3*&6-311++G(3df,3pd)&23.9&7.6&$-7.2$&$5.4$ \\
SCAN+CP&6-311++G(3df,3pd)&\textbf{16.0}&4.9&$2.9$&$5.2$ \\
SCAN+D3+CP&6-311++G(3df,3pd)&19.0&5.2&$-4.9$&\textbf{3.9} \\
SCAN+D3*+CP&6-311++G(3df,3pd)&17.7&\textbf{2.9}&$\mathbf{-1.9}$&$4.1$ \\
\midrule
B2PLYP&Def2-TZVPP&23.1&10.2&$4.8$&$10.3$ \\
B2PLYP+D3&Def2-TZVPP&36.1&11.9&$-11.9$&$7.5$ \\
B2PLYP+CP&Def2-TZVPP&25.2&16.3&$16.3$&$5.1$ \\
B2PLYP+D3+CP&Def2-TZVPP&\textbf{7.7}&\textbf{2.1}&$\mathbf{-0.5}$&\textbf{2.7} \\
\bottomrule
  \end{tabular}
  \caption{The performance of the studied functionals against the reference method. Bold characters mark the smallest MAXD, MAD, MD and SDEP values for each functional.}
  \label{tab:deviats}
\end{table}
\subsection*{\sffamily \large Detailed evaluation of the SCAN results}
As can be seen in Table~\ref{tab:deviats}, that while CP and D3 corrected SCAN does have a smaller standard error of prediction than uncorrected SCAN, its results are, surprisingly, slightly less accurate. In other words, CP and D3 corrected SCAN is more precise, but interaction energies are systematically biased to be more negative than the corresponding reference values as described by the relatively large negative mean deviation of \SI{-4.9}{kJ\per\mole}.

It is possible that D3 over-corrects the SCAN interaction energies. This effect might originate from the fact that SCAN is a meta-GGA functional that implicitly includes mid-range dispersion interactions. Here we used SCAN D3 parameters taken from the Ref. \citenum{Brandenburg2016trw}. However, the authors of that publication optimised only a single D3 parameter, $\alpha_1$. Therefore, we rescaled damping parameters $\alpha_1$ and $\alpha_2$, which control how the dispersion interaction decays over distance (see section 1.8 of the SI for details). The rescaled D3* along with CP-correction eliminates the systematic shift and produces significantly smaller deviations as can be seen in Table~\ref{tab:deviats}.

\begin{figure}[htbp]
\begin{center}
  \includegraphics[width=3.2in]{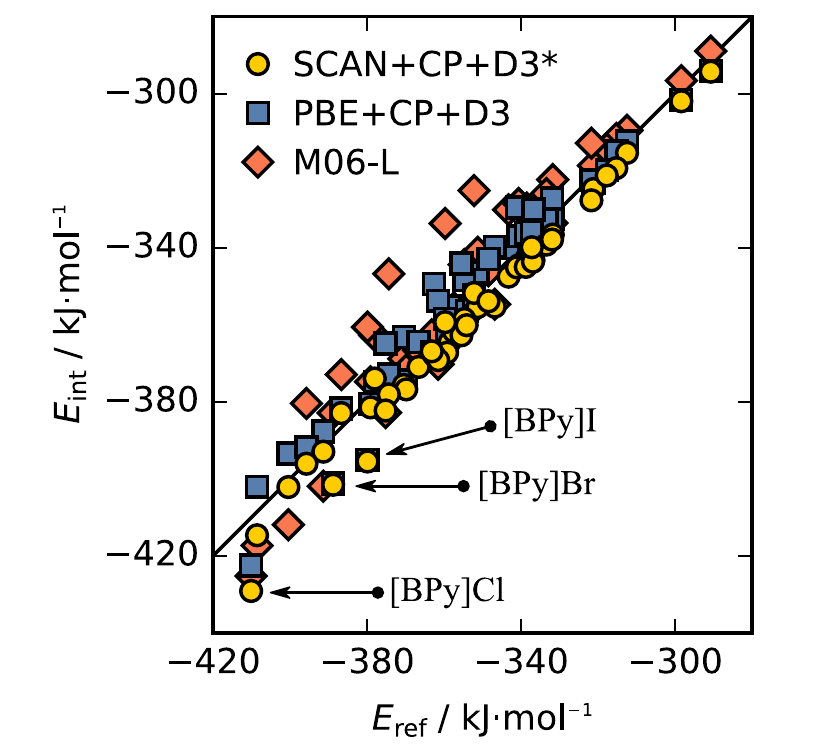}
  \caption{The performance of the tested functionals relative to the reference method.}
  \label{fig:SCANQQ}
  \end{center}
\end{figure}

In Figure~\ref{fig:SCANQQ} the SCAN+CP+D3* interaction energies for each pair of ions are plotted against their corresponding reference values; PBE+CP+D3 and M06-L data points are added for contrast. A similar graph with distinguishable data-points for individual ion pairs calculated with SCAN+CP+D3* is given in the SI (Figure S3). While the overall agreement of the corrected SCAN and the reference method is good, it can be seen that larger deviations occur for some of the more strongly interacting ion pairs. Note that the same outliers also appear for PBE+CP+D3 and M06-L. 

\begin{figure}[htbp]
\begin{center}
  \includegraphics[width=3.25in]{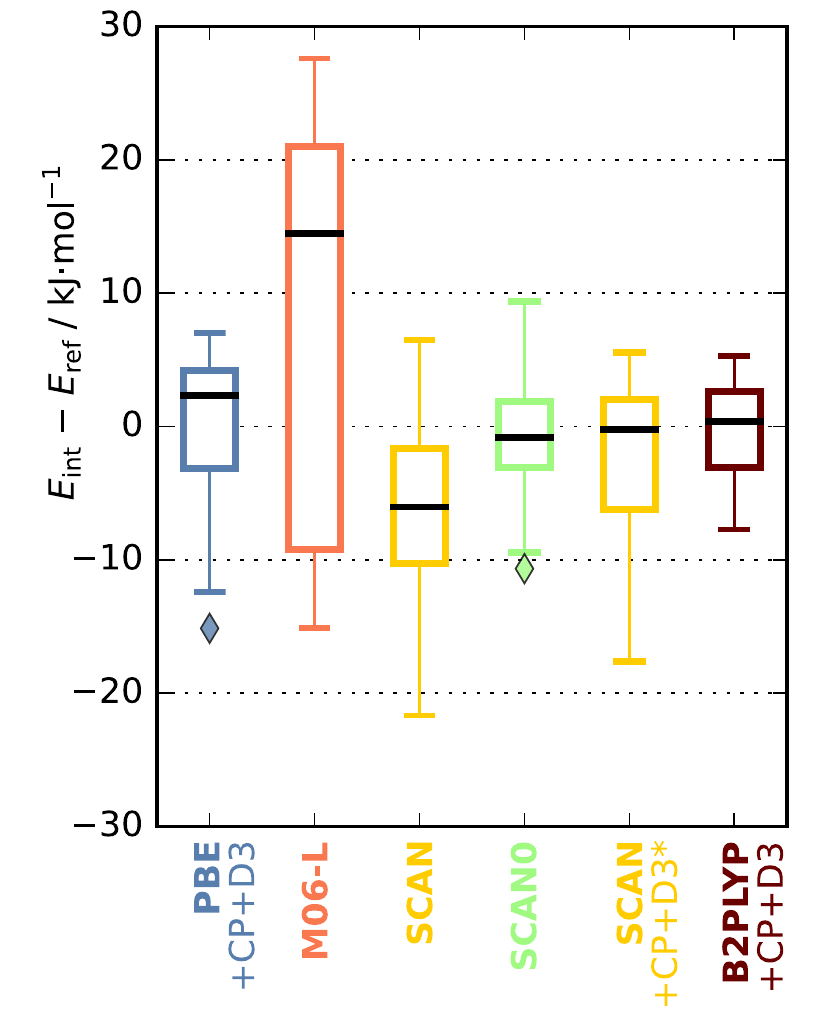}
  \caption{The distribution of errors in interaction energies (relative to the reference method) for the employed DFT methods in describing 12 ion pairs containing chloride, bromide, and iodide anions.}
  \label{fig:hal}
  \end{center}
\end{figure}

The three biggest outliers for the SCAN+CP+D3* method are all 1-butylpyridinium halides (chloride, bromide, and iodide) as specified in Figure~\ref{fig:SCANQQ}. This likely occurs due to the self-interaction error -- an interaction of an electron with itself that is present in approximate DFT methods.\cite{Perdew1981wuv} To avoid this error, we have also calculated the halide anion-containing ion pairs with SCAN0.\cite{Hui2016ust} This hybrid functional includes a Hartree--Fock contribution that negates the effects of the self-interaction error. In ionic associates, the self-interaction error leads to an artificial increase of partial charge transfer between the ions.\cite{Lage-estebanez2016tfb} That is why it is most severe for ion pairs with 1-butylpyridinium cation, as those ion pairs have the largest partial charge transfer due to the proximity of HOMO and LUMO.\cite{Karu2016tki} Increased partial charge transfer directly leads to overestimated interaction energies and underestimated dipole moments. However, application of SCAN0 for the halide anion-containing ion pairs improves the results. This can be judged by examining the distribution of obtained errors in interaction energies shown in Figure~\ref{fig:hal}. Note that due to D3 being unparametrised for SCAN0, no D3 correction was added to the respective energies. It can be seen that compared to SCAN or SCAN+CP+D3*, SCAN0 more accurately predicts the interaction energies for halide anion-containing ion pairs. We conclude that the hybridisation with the exact exchange or self-interaction correction is necessary for systems where there is a large extent of partial charge transfer.
 
\begin{figure}[htbp]
\begin{center}
  \includegraphics[width=3.25in]{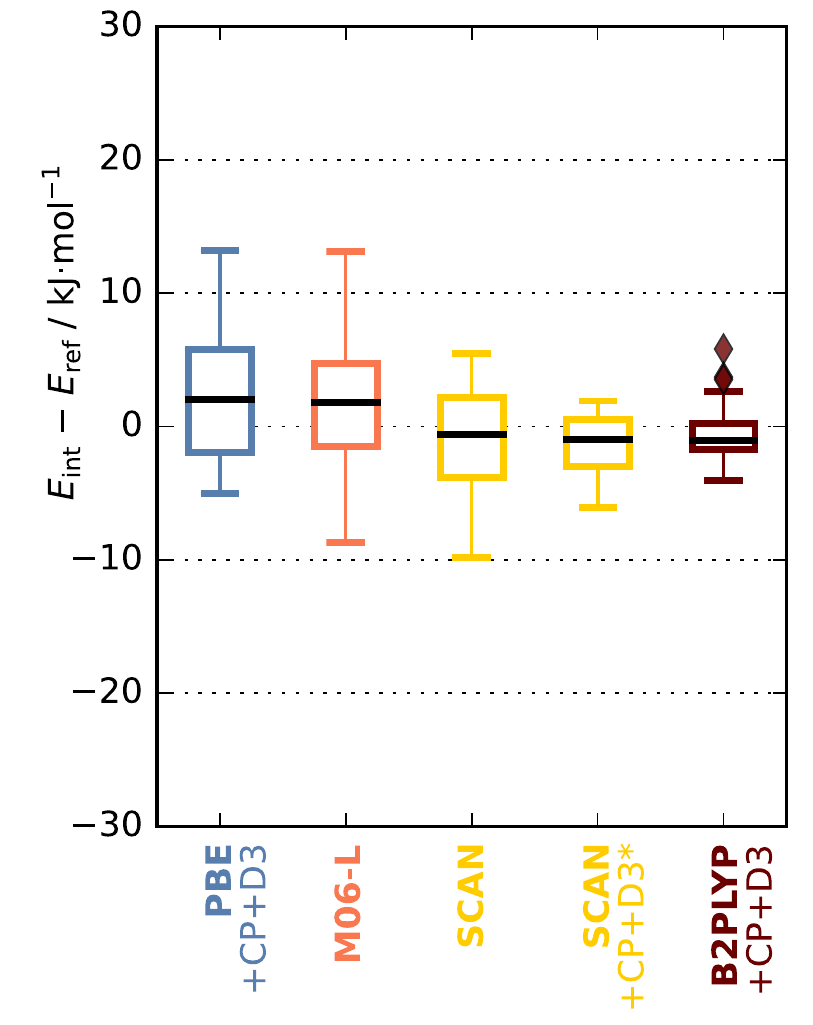}
  \caption{The distribution of errors in interaction energies (relative to the reference method) for the employed DFT methods in describing 36 ion pairs excluding the halide anion-containing ion pairs.}
  \label{fig:nohal}
  \end{center}
\end{figure}

For comparison, in Figure~\ref{fig:nohal} we displayed the performance of the functionals for all ion pairs that do not contain halide anions. The results are similar to those seen in Figure~\ref{fig:deviations}, but with higher accuracy and without the outliers. These findings suggest that ionic liquids excluding chlorides, bromides, and iodides can be effectively studied using the SCAN density functional.

\subsection*{\sffamily \large The comparison of the description of dipole moments}

The charge transfer gives a significant contribution to the dipole moment of ion pairs. Table~\ref{tab:dipoles} demonstrates a comparison of calculated dipole moments. Note that neither CP nor D3 used corrections affect the electronic structure. For this reason, they are omitted. The deviations are calculated against the DLPNO-CCSD(T)/CBS reference method. All predictions are accurate, with $r^2$ even for the worst method exceeding $0.99$. The table suggests that while SCAN functional predicts magnitudes of dipole moments better than PBE, it gives slightly larger errors compared to M06-L and B2PLYP.  

Thus, comparison of the results obtained using the SCAN and M06-L functionals presents an interesting contradiction: SCAN produces a greater error in the calculation of dipole moments but more accurate interaction energies. This indicates that M06-L describes charge transfer better than SCAN but incorrectly estimates its energetic effect. The relatively good performance of M06-L even in comparison with hybrid functionals was seen in the work of Lage-Estebanez \textit{et al.},\cite{Lage-estebanez2017vki} where authors related the charge transfer to the self-interaction error. So, we concluded that the self-interaction error correction could be used to improve the overall SCAN performance.

Earlier Perdew suggested that there are two roads to follow to alleviate the self-interaction error.\cite{Perdew2015uub} One is to apply the Perdew--Zunger self-interaction error correction to a pure DFT functional,\cite{Perdew1981wuv} and the other one is to use hybrid functionals, such as SCAN0. In this work, we tested the second road, leaving the first one for a separate study. It can be seen in Table~\ref{tab:dipoles0} that SCAN0 surpasses all of the studied non-hybrid functionals in accuracy and rivals the double-hybrid B2PLYP.

\begin{table}[!htbp]
  \centering
\begin{tabular}{l|rrr|r}
\toprule
  Error &    PBE &   SCAN &    M06-L &  B2PLYP \\
\midrule
     MAXD  & $-0.505$ & $-0.381$ & $\mathbf{-0.358}$ & $-0.108$ \\
      MAD  &  0.216 &  0.131 &  $\mathbf{0.110}$ &   0.043 \\
      MD   & $-0.216$ & $-0.131$ & $\mathbf{-0.110}$ &  $-0.036$ \\
      SDEP &  $0.087$ &  $0.075$ &  $\mathbf{0.064}$ &  $0.038$ \\
  ${r^2}$ &  0.993 &  $0.994$ &  $0.995$ &   $0.998$ \\
  \bottomrule
  \end{tabular}
  \caption{The deviations of the magnitude of dipole moments in Debye for the studied functionals. The smallest MAXD, MAD, MD and SDEP values among the non-hybrid functionals are marked by bold.}
  \label{tab:dipoles}
\end{table}

\begin{table}[!htbp]
  \centering
\begin{tabular}{l|rrrr|r}
\toprule
  Error &    PBE &   SCAN & SCAN0 &     M06-L &  B2PLYP \\
\midrule
MAXD & $ -0.489 $ & $ -0.325 $ & $\mathbf{-0.129} $ & $ -0.282 $ & $ 0.109 $ \\
MAD & $ 0.331 $ & $ 0.120 $ & $ \mathbf{0.056} $ & $ 0.150 $ & $ 0.033 $ \\
MD & $ -0.331 $ & $ -0.120 $ & $ \mathbf{-0.015} $ & $ -0.150 $ & $ 0.021 $ \\
SDEP & $ 0.077 $ & $ 0.099 $ & $ \mathbf{0.068} $ & $ 0.069 $ & $ 0.038 $ \\
${r^2}$ & $ 0.996 $ & $ 0.995 $ & $0.996 $ & $ 0.996 $ & $ 0.998 $ \\
  \bottomrule
  \end{tabular}
  \caption{The deviations of the magnitude of dipole moments in Debye for the studied functionals when describing halide containing ion pairs. The smallest MAXD, MAD, MD and SDEP values among the non-hybrid functionals and SCAN0 are marked by bold.}
  \label{tab:dipoles0}
\end{table}










\section*{\sffamily \Large CONCLUSIONS}

In this work, we have evaluated the performance of the recently proposed strongly constrained and appropriately normed (SCAN) density functional on a dataset of 48 ionic liquid ion pairs. The main focus was on the interaction energies and dipole moments; their predictions with SCAN were compared to the values of DLPNO-CCSD(T) and DFT methods. Our key findings are the following:
\begin{itemize}

\item SCAN is a fast and accurate method for evaluating interaction energetics of ionic liquid associates. Therefore, SCAN is expected to be suitable for both high-throughput screenings as well as DFT-based molecular dynamics simulations of ionic liquids.
\item More accurate interaction energies for SCAN are obtained in combination with Grimme's D3 dispersion correction and Boys--Bernardi counterpoise corrections. The mean absolute deviation of the corrected SCAN functional for our dataset is \SI{5.2}{kJ\per\mole}, yet the deviation is clearly systematic, which leads us to believe that it could be taken into account. We show that by tuning the D3 parameters, the deviation mean absolute deviation can be reduced to below \SI{2.9}{kJ\per\mole}.
\item SCAN is sensitive to the self-interaction error, which is demonstrated in the example of ion pairs containing chloride, bromide, and iodide anions.
\end{itemize}


\subsection*{\sffamily \large ACKNOWLEDGMENTS}


The EU supported this research through the European Regional Development Fund (Centre of Excellence, 2014-2020.4.01.15-0011), Institutional Research Grant IUT20-13, and Estonian Personal Research Project PUT1107. Results were obtained in part using the High Performance Computing Center of the University of Tartu and in part using the EPSRC funded ARCHIE-WeSt High Performance Computer (www.archie-west.ac.uk, EPSRC grant no. EP/K000586/1). JS thanks the support from the Center for the Computational Design of Functional Layered Materials, an Energy Frontier Research Center funded by the US Department of Energy (DOE), Office of Science, Basic Energy Sciences (BES), under award no. DE-SC0012575.

\noindent The authors declare no competing financial interest.

\noindent Additional Supporting Information may be found in the online version of this article.

\clearpage


\bibliography{main.bbl}   


\end{document}